\documentclass[aps,twocolumn,superscriptaddress,amsmath,amssymb]{revtex4}

\usepackage{graphicx}
\usepackage{xcolor}

\begin{document}

\title{Inheritances, social classes, and wealth distribution}

\author{P. Patr\'{\i}cio}
\email{pedro.patricio@isel.pt}
\affiliation{Departamento de F\'{\i}sica, ISEL - Instituto Superior de Engenharia de Lisboa, 
Instituto Polit\'ecnico de Lisboa, 1959-007 Lisboa, Portugal}
\affiliation{Centro de F\'{\i}sica Te\'orica e Computacional, Faculdade de
Ci\^{e}ncias, Universidade de Lisboa, 1749-016 Lisboa, Portugal}

\author{N. A. M. Ara\'ujo}
\affiliation{Departamento de F\'{\i}sica, Faculdade de Ci\^{e}ncias, Universidade de Lisboa, 1749-016 Lisboa, Portugal}
\affiliation{Centro de F\'{\i}sica Te\'orica e Computacional, Faculdade de
Ci\^{e}ncias, Universidade de Lisboa, 1749-016 Lisboa, Portugal}

\begin{abstract}
We consider a simple theoretical model to investigate the impact of
inheritances on the wealth distribution. Wealth is described as a finite
resource, which remains constant over different generations and is divided
equally among offspring. All other sources of wealth are neglected. We
consider different societies characterized by a different offspring
probability distribution. We find that, if the population remains constant,
the society reaches a stationary wealth distribution. We show that inequality
emerges every time the number of children per family is not always the same.
For realistic offspring distributions from developed countries, the model
predicts a Gini coefficient of $G\approx 0.3$. If we divide the society into
wealth classes and set the probability of getting married to depend on the
distance between classes, the stationary wealth distribution crosses over from
an exponential to a power-law regime as the number of wealth classes and the
level of class distinction increase.
\end{abstract}

\maketitle

\section{Introduction}

Empirical wealth distributions are characterized by two enduring features.
For the large majority of the population, which has small or medium wealth
$w$, the distribution is positively skewed, roughly resembling a lognormal
distribution.  However, the tail for the wealthier is well approximated by a
power-law distribution \cite{pareto1897cours}:
\begin{equation} \label{eq:paretoslaw}
f(w)\propto\frac{1}{w^\alpha},
\end{equation}
also referred to as a Pareto law. Although this law refers only to the
wealthier and, therefore, to a small percentage of the population, its
importance may not be overlooked, as it concerns the richest part of the
population, holding the larger percentage of the total wealth.  The more
unequal the society is, the smaller is the value of $\alpha$.  The data
regarding labor income is now very well documented, and the corresponding
$\alpha$ varies between 1.5 and 3 \cite{clementi2005pareto,atkinson2011top}.  
The past forty years have seen a
disturbing increase in income inequality (and consequently smaller values of
$\alpha$) almost everywhere in the world \cite{piketty2019capital}.  General
wealth distributions are difficult to find, as they concern all material
assets, in the form of real property and financial claims. Nevertheless,
almost all studies find that the wealth distribution is more unequal than the
labor income distribution \cite{davies2000distribution}.

The ubiquitous Pareto law, which also appears in other socio-economic contexts, 
such as firm or city sizes \cite{ball2004critical,gabaix2009power}, hints at
some universality, which should be robust to the fine details of the theoretical 
model we use to describe a society.
Many models have been proposed to explain the tail distribution of wealth 
\cite{champernowne1953model,simon1955class,wold1957model,mandelbrot1961stable} 
(or, more recently \cite{levy1996power,bouchaud2000wealth}, in the context of physics),
mainly along the lines of random growth, which assumes Gibrat's law of proportionate effect.
This law states that the distribution of the percentage
growth rate of a unit (e.g. wealth, the size of a firm or a city) is independent of its size.

If one aims at an understanding of the forces that contribute to larger or
smaller wealth inequalities, the explicit mechanisms behind wealth inequality
must be incorporated. This rapidly leads to complex models that are difficult
to analyze. Indeed, the reasons behind wealth inequality are innumerable: we
have, first of all, the inheritance and education we receive from our parents,
the marriages or alliances we make, associated so many times with the
relatively closed circles of relationships we establish, our business talent
and ability to work, our age and health or simply mere luck. This article does
not intend to make an extensive literature review about these economic models.
The interested reader is referred to Refs.
\cite{davies2000distribution,de2017saving}, for a more economical perspective
and Ref. \cite{yakovenko2009colloquium}, for a more physical one.

For the sake of simplification, these models may be divided into two types.
Lifecycle models (LCM) consider the wealth evolution during an individual
lifetime, in which inheritances play no role. These are also known as
intragenerational models. Other models suppress interest in lifecycle
variations and focus on intergenerational links. Very few contributions have
attempted to deal simultaneously with both the lifecycle and inherited
components of wealth \cite{davies2000distribution}.

The simple model proposed in our article in the context of statistical physics
belongs to the second type: we intend to quantify the evolution of the
distribution of wealth over several successive generations.  For each cohort,
the sum of the wealth of all individuals is considered to be constant.  In our
model, wealth could be thought as a finite resource of the society, which
remains constant and must be divided by all individuals.  Of all possibilities
for enrichment, we will focus on two particular main aspects.  On the one
hand, the variable number of children of each family, which implies different
inheritances.  If we assume that the inheritance is equally divided by all
children, the smaller the number of children, the greater the inheritance of
each child. On the other hand, the fact that people tend to marry people with
comparable wealth, or belonging to the same social circle or class. These two
factors will inevitably lead to an unequal distribution of wealth, even if we
start from a very homogeneous society.

One of the first intergenerational models \cite{blinder1973model} is closely
related to the model we present here.  It considers a simplified society in
which every family has exactly two children, a boy and a girl.  This model
discusses the implications for wealth inequality of primogeniture, when all
the family fortune is given to the male heir, equal division, or unequal
division.  It also discusses the effect of having random mating, in which
there is no relation between the wealth of the husband and the wife, class
mating, in which the wife has exactly the same wealth of the husband, or an
assortative mating, something in between.  However, this model never discusses
other offspring distributions, as we do in our article.  Other
intergenerational models considered societies with individuals of different
age, with a mortality probability distribution \cite{gokhale2001simulating},
or other complex features regarding personal earnings, consumption, savings
and motives of bequests \cite{becker1979equilibrium,de2017saving}.

This paper is organized as follows. In section \ref{sec_model}, we present our
intergenerational model, which is characterized by a particular marriage and
offspring probability distributions. In section \ref{sec_results}, we describe
the results we obtain for a society without and with well defined classes.
Particular emphasis is given to the stationary wealth distributions we
determine in each case.  In the last section, we discuss the importance of our
results both in the context of economic inequality and statistical physics.

\section{Model}
\label{sec_model}

We consider a society composed of $N_0$ individuals with an initial
distribution of wealth and gender. For individual $i$ ($i=1,...,N_0$), the
wealth $w_i$ is drawn from an initial {\sl wealth} probability distribution
$f_0^w(w)$. The gender $g_i$ is either ``female'' or ``male'', with equal
probability. 

We consider that marriages among people with comparable wealth are more
likely.  Thus, we organize the society into $N_c$ classes with $N_0/N_c$
individuals each. Individuals are organized into classes following the rank of
increasing wealth. For simplicity, we only consider different-gender
marriages.  If $i$ and $j$ are of a different gender, the probability of
getting married $f^m_{ij}$ is
\begin{equation}
f^m_{ij}\propto e^{-\beta d_{ij}}, \label{eq_mpd}
\end{equation}
where $d_{ij}$ is defined as a distance between their classes and $\beta$ is
the level of class distinction, or the inverse of the level of mixing between
classes. For $\beta=0$, $f^m_{ij}$ is the same for all pairs and there is no
distinction between classes. The larger the value of $\beta$ is, the more
likely it is that marriage between individuals in the same class are favored
over inter-class marriage.  Here, we consider $d_{ij}=|c_j-c_i|$, where $c_i$
and $c_j$ is the rank of the class, when they are all ordered by increasing
wealth.

We select the pairs to couple in the following way. For each individual $i$,
we randomly select from which class the couple $j$ is, where the probability
$p(c_j)$ for each class $c_j$ is,
\begin{equation}
p(c_j)=\frac{n(c_j)f^m_{ij}}{\sum_{c_k}{n(c_k)f^m_{ij}}}.
\end{equation}
$n(c_j)$ is the number of individuals in class $c_j$ that are of a different
gender than $i$ and the sum is over all classes. We then randomly select one
individual $j$ to marry $i$ among the $n(c_j)$.  Note that, instead of using
classes, we could think of a marriage probability distribution that depends
directly on the wealth difference $d_{ij}=|w_j-w_i|$.  However, this
methodology, which is equivalent in the limit $N_c\to N_0$, is computationally
much more demanding.

Once all couples are defined, a new generation of individuals is generated.
Each married couple $ij$ is replaced by $o_{ij}$ offspring according to a
specific {\sl offspring} probability distribution $f^o(o)$, and the total
wealth of the parents is equally distributed among the offspring.  Each
individual of the new generation is either a ``female'' or a ``male'' with
equal probability.  Here, we represent the complete offspring discrete value
distribution by the set
\begin{equation}
f^o=\{f^o(0),f^o(1),...,f^o(n_\mathrm{max})\},\label{eq_opd}
\end{equation}
where $n_\mathrm{max}$ is the maximum number of offspring per family.

Due to the stochastic nature of the dynamics, for each generation, the number
of ``female'' and ``male'' individuals is only equal on average. Thus, at the
end of the matching protocol some excess individuals of a given gender will be
unpaired or without offspring. We redistribute their wealth equally among all
individuals of the new generation. So, the total wealth is conserved at all
times.

Once a new society with $N_1$ individuals is formed, classes are redefined
according to the new wealth distribution $f_1^w(w)$, and the process of
generating the next generation is repeated.

\section{Results}
\label{sec_results}

\subsection{Societies without classes}
Let us first consider a society of $N_0=10^5$ without classes, where all pairs
of individuals of a different gender are equally likely to get married, i.e.,
$N_c=1$ or $\beta=0$. We set 
\begin{equation}
f^o=\{0,0,1\} \ \ ,
\end{equation}
which corresponds to exactly two offspring per couple and the size of the
society remains approximately constant. To characterize the level of wealth
inequality, we compute the Gini coefficient $G$, defined as,
\begin{equation}
G=1-\frac{2}{N^2}\sum_{i=1}^N \sum_{j=1}^i \frac{w_j}{\mu} 
\end{equation}
where $N$ is the size of the population, $\mu$ the average wealth and the sum
follows the rank of increasing wealth. $G=0$ for an egalitarian society, where
$w_i$ is the same for all individuals, and $G\approx 1$ for a large society
where all the wealth is concentrated in a few number of individuals. 

\begin{figure}
\begin{center}
\includegraphics[width=\columnwidth]{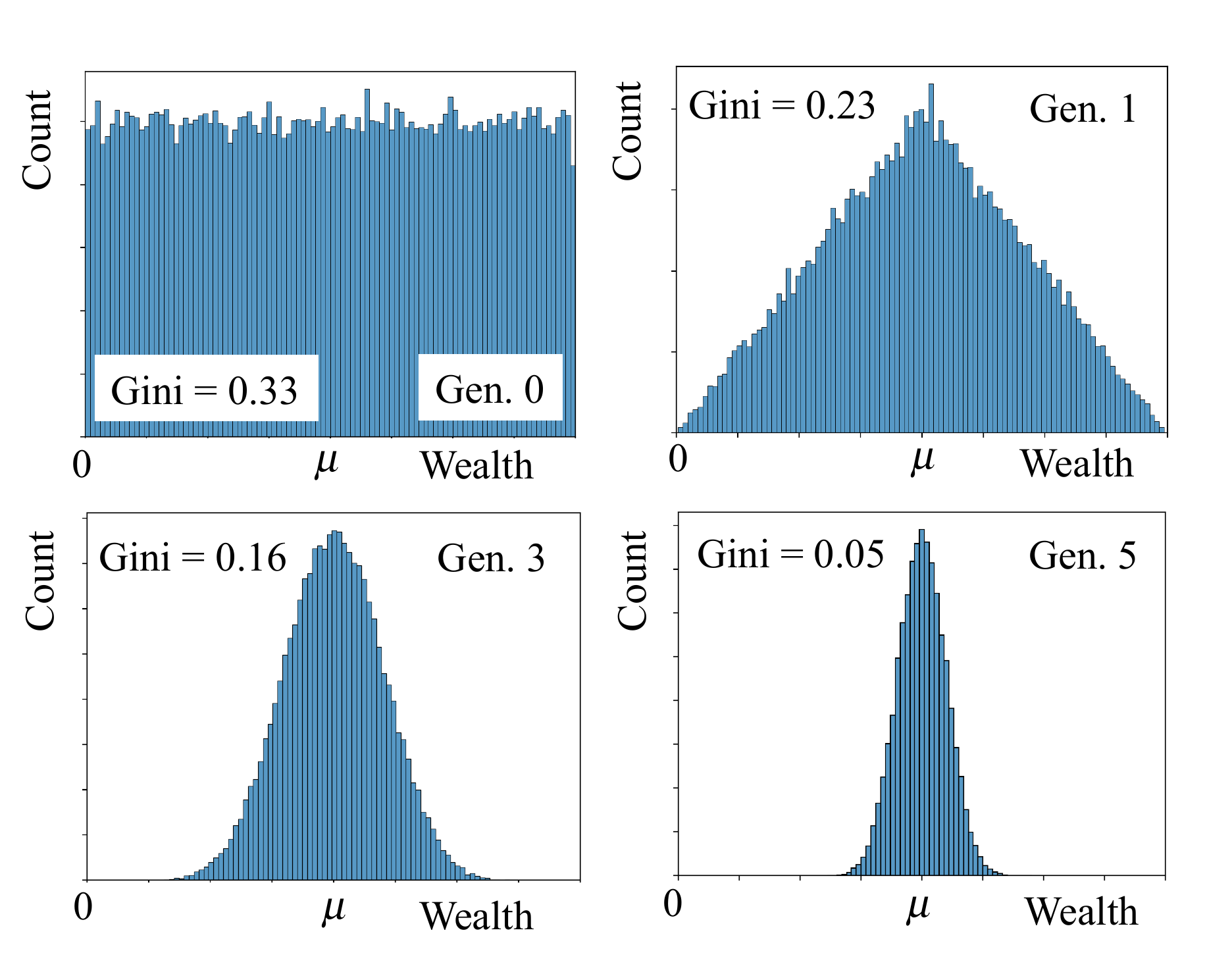}
\caption{Evolution of the wealth distribution for a society without classes
and exactly two offspring per couple. Initially, the wealth distribution is
uniform, $f_0^w(w_i)=1/(2\mu)$, for $0<w<2\mu$ (left-upper panel). In the
first generation the wealth distribution is triangular (right-upper panel).
For later generations $g$, the wealth distribution converges to a Gaussian
distribution, with average $\mu$ and variance $\sigma_g^2=\sigma_0^2/2^g$,
where $\sigma_0$ is the initial variance. Thus, the Gini coefficient vanishes
exponentially.~\label{Fig1}}
\end{center}
\end{figure}
We set the initial distribution of wealth to be uniform, of average $\mu$,
with $f_0^w(w_i)=1/(2\mu)$ and $0<w_i<2\mu$, which corresponds to $G=1/3$.
Figure~\ref{Fig1} shows the wealth distribution for four different
generations. The Gini coefficient rapidly converges to zero. This is in fact
the case for any initial wealth distribution. Since individuals are paired at
random and their wealth evenly distributed among their two offspring, at each
iteration pairwise heterogeneities in the wealth distribution are reduced.
Precisely, the average wealth remains $\mu$ at all times, but the variance in
generation $g$ is $\sigma_g^2=\sigma_0^2/2^g$, which vanishes asymptotically.
So, for any initial distribution of wealth, the society rapidly converges
towards an egalitarian society where all individuals have (approximately) the
same wealth. This result was also obtained by Blinder \cite{blinder1973model},
in his intergenerational model in which each family had two children, a boy
and a girl.

In fact, equality emerges for any society in which each couple has exactly the
same number of children $n$. If $n<2$ ($n>2$), the population decreases
(increases) exponentially.  However, the result
$(\sigma_g/\mu_g)^2=(\sigma_0/\mu_0)^2/2^g$ still holds.

\begin{figure}
\begin{center}
\includegraphics[width=\columnwidth]{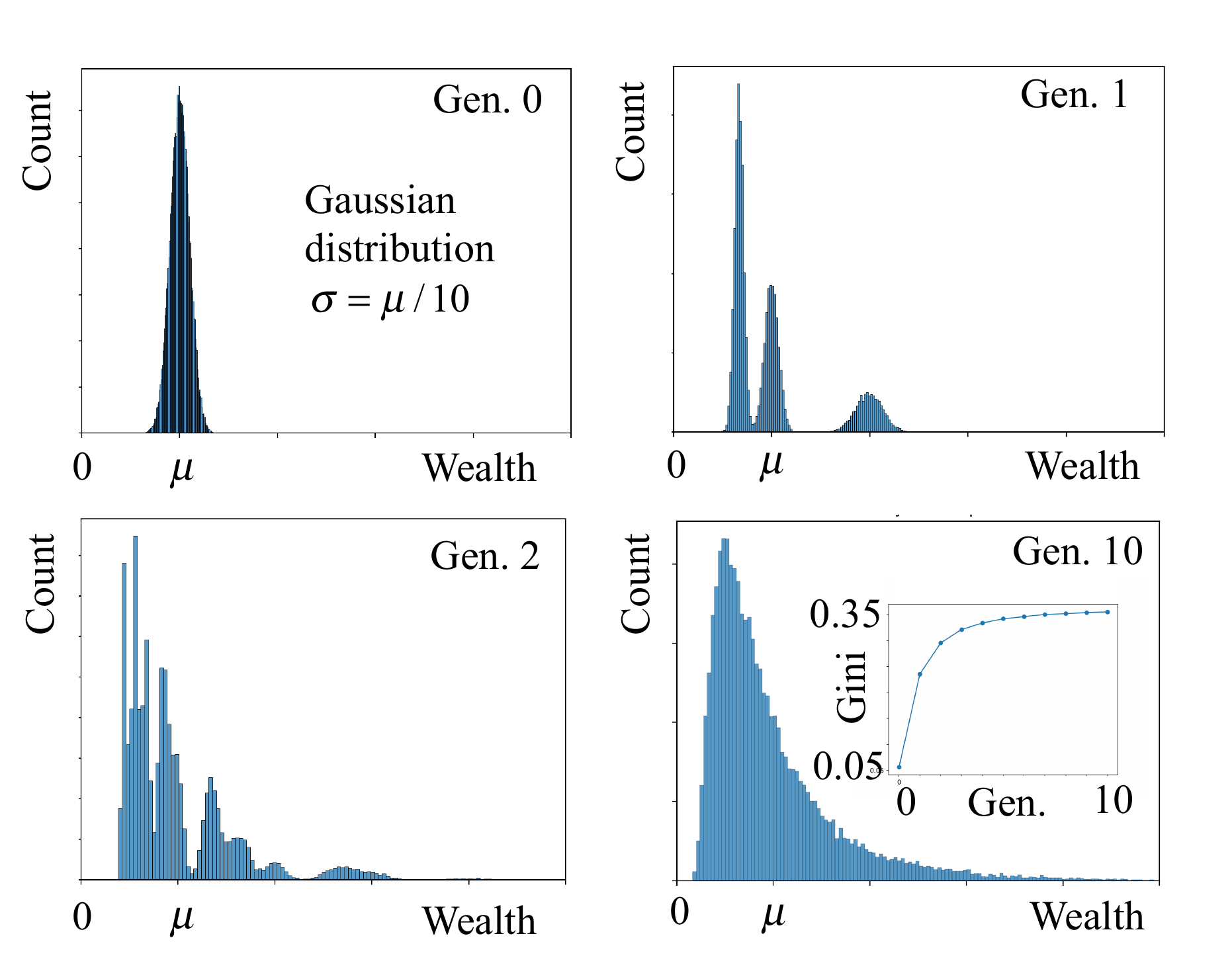}
\caption{Evolution of the wealth distribution for a society without classes
and $f^c=\{0,1/3,1/3,1/3\}$. Initially, the wealth distribution is a Gaussian
of average $\mu$ and $\sigma=\mu/10$ (left-upper panel). In the first
generation (right-upper panel), the distribution is a sequence of three peaks,
which correspond to individuals from families with one, two, and three
offspring. In the second generation (left-bottom panel), each peak is split
into three.  The wealth distribution converges rapidly to a broad stationary
distribution, as shown in the right-bottom panel for generation ten. In the
inset is the evolution of the Gini coefficient with the generation, which in
three generations goes from $\approx 0.05$ to $\approx 0.35$.  \label{Fig2}}
\end{center}
\end{figure}
\begin{figure}
\begin{center}
\includegraphics[scale=0.5]{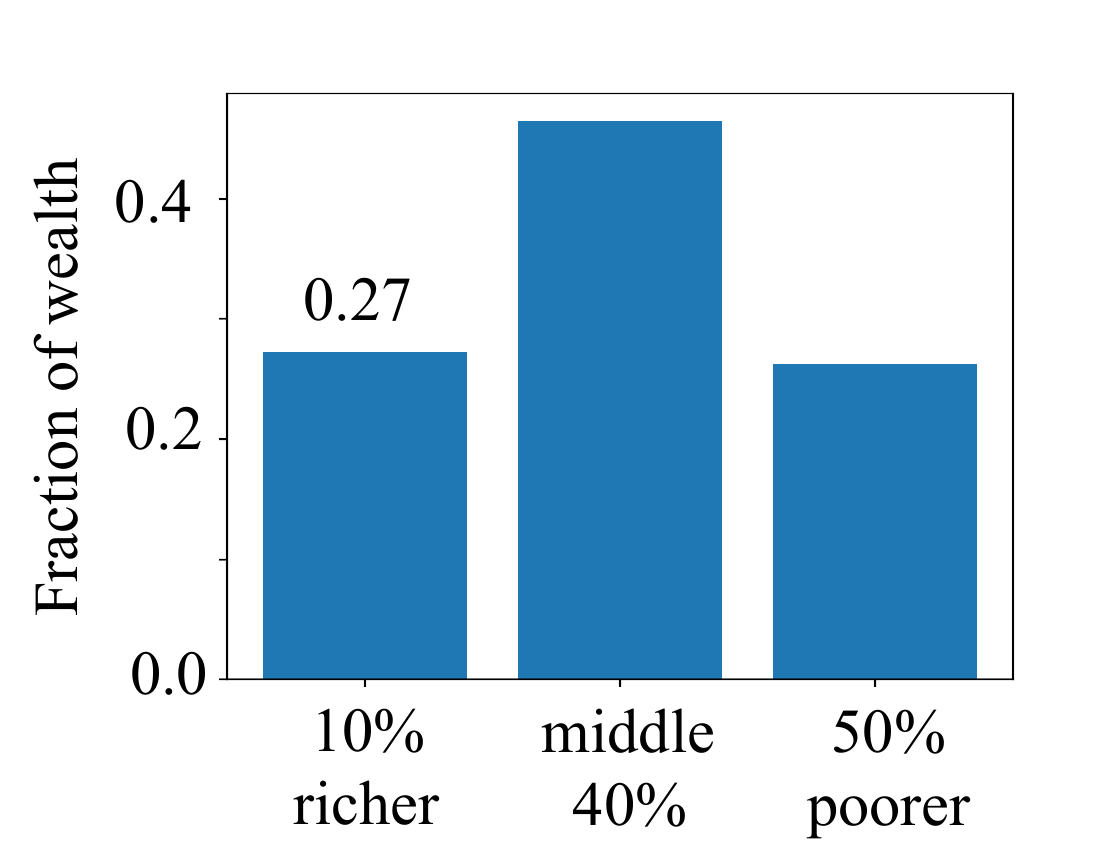}
\caption{Distribution of the wealth among three different groups: $10\%$
richer, the middle $40\%$, and the $50\%$ poorer. Results are for generation
ten in the society of Fig.~\ref{Fig2}, which are a good approximation of the
stationary distribution.~\label{Fig3}}
\end{center}
\end{figure}

We now consider a society with the same $N_0$ but a different offspring
probability distribution,
\begin{equation}
f^o=\{0,1/3,1/3,1/3\} \ \ .
\end{equation}
Note that, since the average offspring per couple is two, the size of the
society remains (approximately) constant. Figure~\ref{Fig2} shows the wealth
distribution for different generations. Starting from a Gaussian distribution
of average $\mu$ and $\sigma=\mu/10$, which yields a Gini coefficient
$G\approx 0.05$. In the first generation, the distribution is characterized by
a sequence of three peaks, which correspond to individuals from families with
one, two, and three offspring. The third peak (higher wealth) has average
$2\mu$ and includes $1/6$ of all individuals, which are the ones from families
with only one offspring. The peak in the middle, is centered at $\mu$ and
accounts for $1/3$ of the population, corresponding to the individuals from
families with two offspring. The peak on the left has average $\mu/3$ and the
narrowest dispersion and corresponds to the $1/2$ of the population, belonging
to families with three offspring. After a few generations, the wealth
distribution rapidly converges to a well-defined distribution, as   shown for
generation ten in the figure. The Gini coefficient increases with the
generation and converges to $G\approx 0.35$ after a few iterations. This
suggests that wealth inequality is observed even for a society without
classes, provided that the number of offspring per family is not always two.
Figure~\ref{Fig3} shows the fraction of wealth distributed between three
different groups: the $10\%$ richer, the middle $40\%$, and the $50\%$ poorer.
The first group has $27\%$ of the total wealth, which is slightly more than
the third group, which means that, on average, an individual of the first
group have five times more wealth than one from the third group.

\begin{figure}
\begin{center}
\includegraphics[width=\columnwidth]{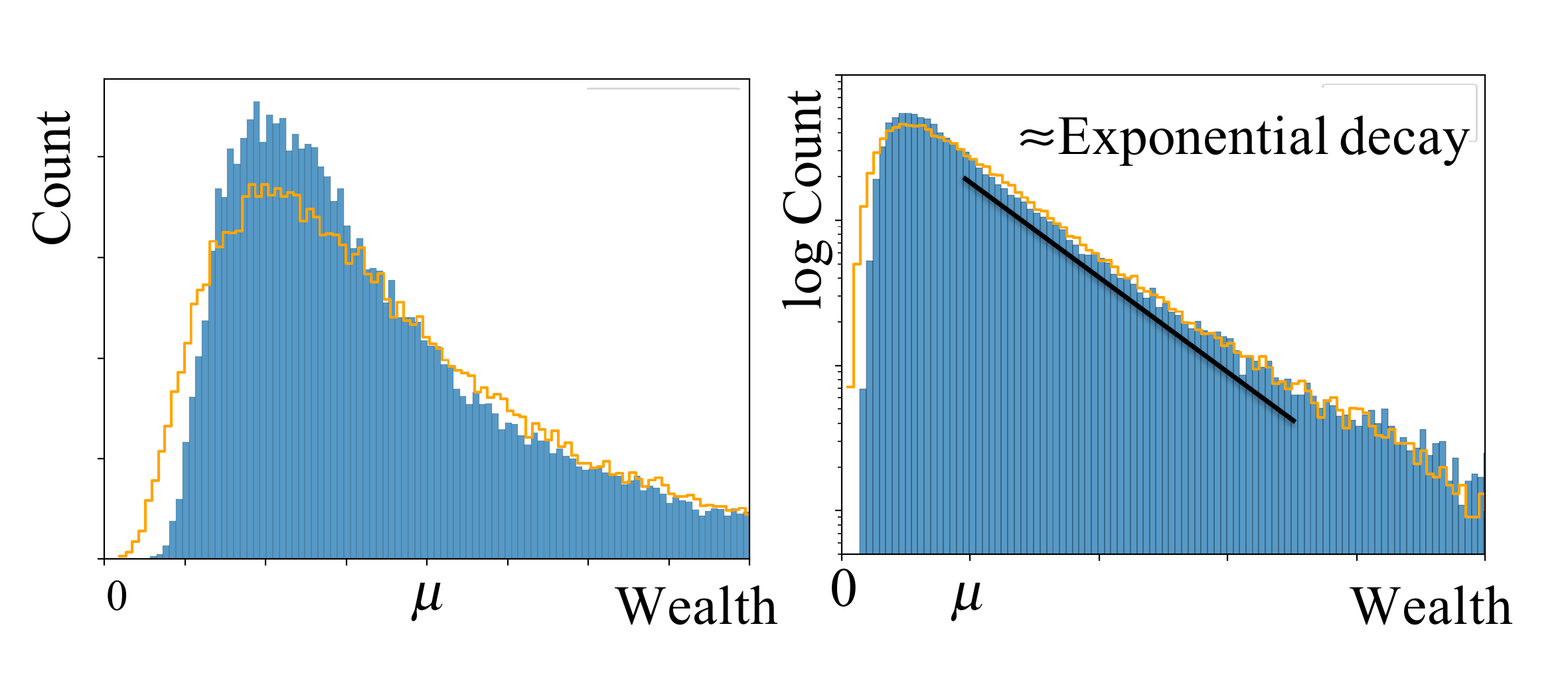}
\caption{Stationary wealth distribution for a society without classes, in a
linear-linear (left) and a log-linear (right) scale.  Results are for
generation ten in the society of Fig.~\ref{Fig2}.  The (orange) solid line
corresponds to a log-normal distribution with the same average wealth and
variance. In the right panel,  the (black) solid line corresponds to an
exponential decay with a characteristic wealth of $w_c\approx
1,6\mu$.~\label{Fig4}}
\end{center}
\end{figure}
The shape of the wealth distribution reported for generation ten is in fact
very robust and corresponds to a stationary distribution.  We simulated higher
generations and found no visual differences between the distributions.  We
have also considered other initial configurations, such as, for example,
uniform and a bimodal distribution and, for all of them we obtained the same
stationary wealth distribution, after a proper rescaling by the average wealth
$\mu$. Figure~\ref{Fig4} shows this stationary wealth distribution in a
linear-linear and a log-linear scale. It is clear that, for large values of
the wealth, the distribution decays logarithmic. For the sake of comparison,
we also represented in a solid line a log-normal distribution with the same
average wealth and variance. The main features of the wealth distribution are
well captured by a log-normal distribution. Notwithstanding, the log-normal
has a larger population for lower values and a slightly lower peak.

We investigated several ``societies'' with other offspring probability
distributions, including more realistic distributions, such as:
\begin{equation}
f^o=\{0.1,0.2,0.4,0.2,0.1\} \ \ .
\end{equation}
For this particular choice, which follows approximately the recent statistical
bulletin~\cite{2020offspringdistribution}, but in which the size of the
society remains constant, the Gini coefficient rapidly converges to
$G\approx0.28$.  We also considered some eccentric offspring distributions,
which led us to stationary wealth distributions with smaller or larger Gini
coefficients.  As a general rule, as the variety of the number of offspring
per family increases, also increases the wealth inequality.

\subsection{Societies with classes}

We now study the impact of having a probability of getting married that
depends on the class of each individual (see section \ref{sec_model}).  For
simplicity, we consider the representative offspring probability distribution
where each couple has equal probability to have one, two, or three offspring,
which corresponds to $f^o=\{0,1/3,1/3,1/3\}$.

Let us consider first a society with $N_c=3$ classes, where the poorer $N_0/3$
individuals are in the lower class, the next $N_0/3$ in the middle one, and
the richer $N_0/3$ individuals in the upper one. As explained in section
\ref{sec_model}, we consider a marriage probability between individuals $i$
and $j$ proportional to $\exp(-\beta d_{ij})$, where $d_{ij}=|c_j-c_i|$ is the
difference between the number of their classes. For $\beta=1$, for a society
with the same number of individuals per class, the probability for an
individual to marry someone from the same class is only $e\approx 2.7$ larger
than the one of marrying with someone from a neighboring class. However, for
$\beta=10$, this factor increases by four orders of magnitude and so it is
practically impossible to have marriages between individuals of different
classes, except when someone in a class is left without a pair. 

\begin{figure}
\begin{center}
\includegraphics[width=\columnwidth]{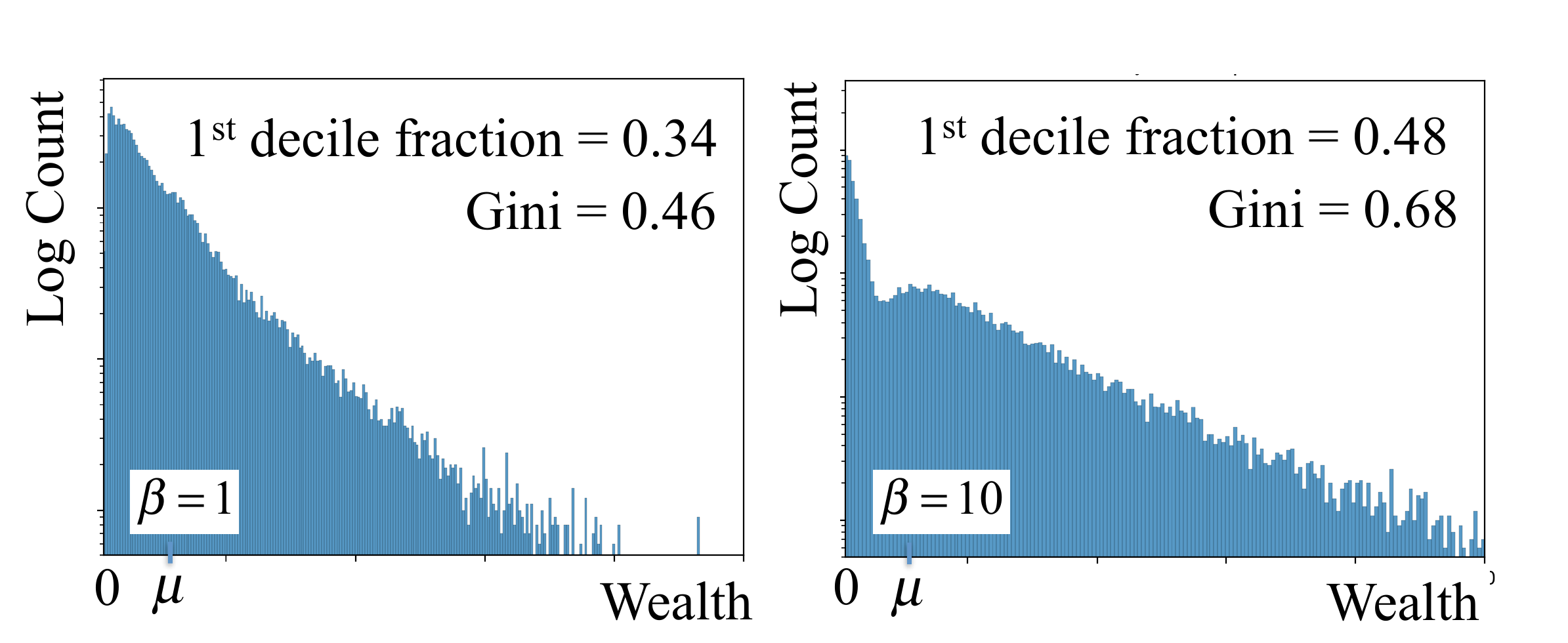}
\caption{Stationary wealth distributions for a society with three classes for
$\beta=1$ (left) and $\beta=10$ (right).  For $\beta=1$, the Gini coefficient
is $0.46$ and the $10\%$ richest individuals accumulate $34\%$ of the total
wealth. For $\beta=10$, the Gini coefficient is $0.48$ and the top $10\%$
accumulate $48\%$ of the total wealth.~\label{Fig5}}
\end{center}
\end{figure}
We start with $N_0=2\times10^5$ and a Gaussian wealth distribution of average
$\mu$ and $\sigma=\mu/10$. As before, the wealth distribution rapidly
converges after a few generations.  Figure~\ref{Fig5} shows the stationary
wealth distribution for $\beta=1$ (obtained at generation 10) and $\beta=10$
(obtained at generation 20). The distribution for the society with higher
degree of mixing between classes ($\beta=1$) is similar to the one found for a
society without classes in Fig.~\ref{Fig3}, but with a higher Gini coefficient
($G\approx0.46$) and with $34\%$ of the wealth concentrated in the top $10\%$
of the population. For the society with lower degree of mixing ($\beta=10$)
the inequalities are even more evident. The Gini coefficient is $G\approx0.68$
and the top $10\%$ accumulate $48\%$ of the total wealth.

\begin{figure}
\begin{center}
\includegraphics[width=\columnwidth]{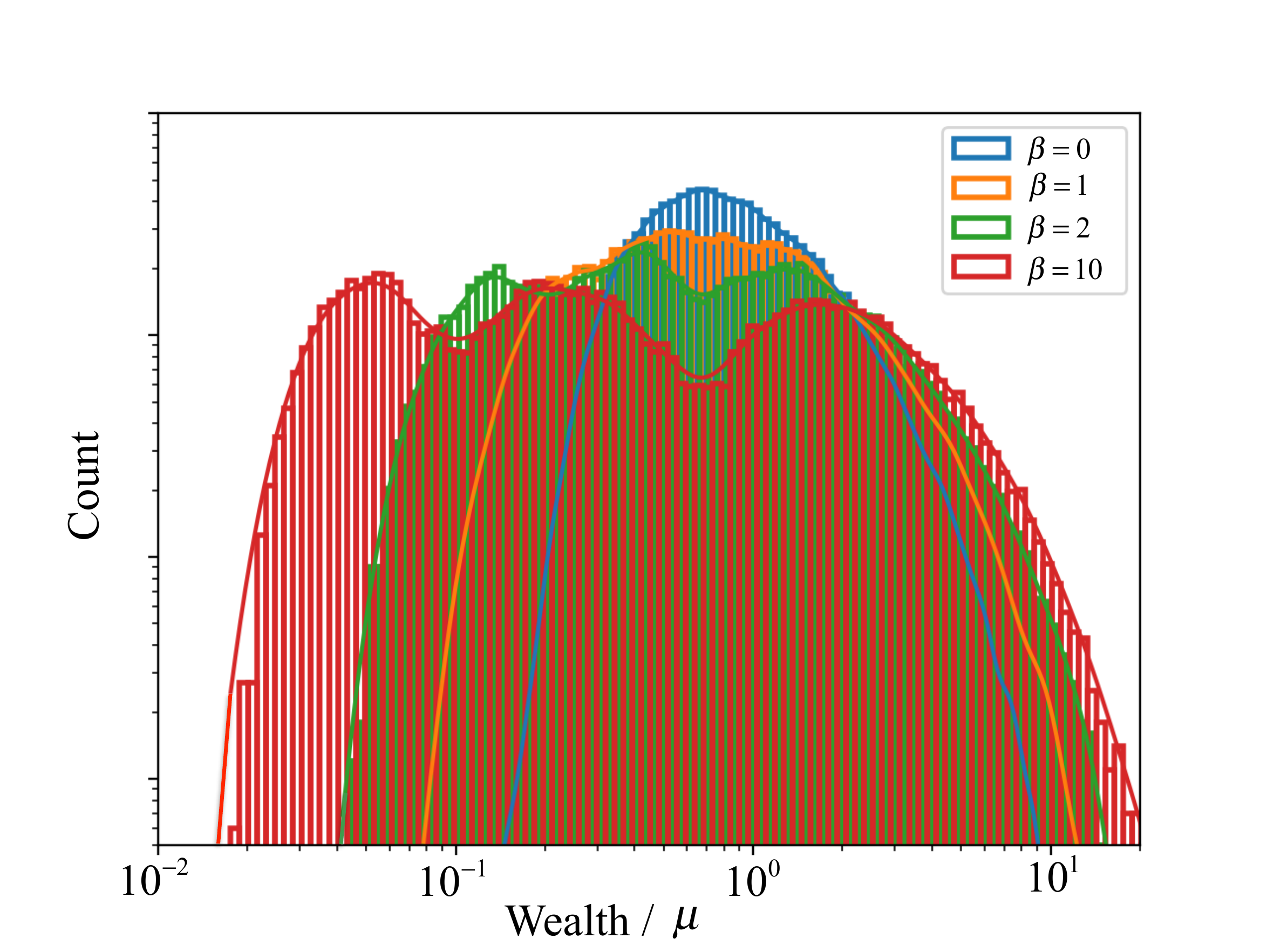}
\caption{Stationary wealth distributions for a society with three classes, for
$\beta=\{0,1,2,10\}$ (blue, orange, green, and red) and
$N_0=2\times10^5$.~\label{Fig6}}
\end{center}
\end{figure}
Figure~\ref{Fig6} shows the stationary wealth distribution for different
values of $\beta$. For $\beta=0$, the wealth distribution shows a well-defined
peak around the average wealth $\mu$, as discussed before. As $\beta$
increases, the degree of mixing between classes is exponentially reduced and
the distribution becomes broader and consisting of a sequence of three
overlapping peaks (one per class).

\begin{figure}
\begin{center}
\includegraphics[width=\columnwidth]{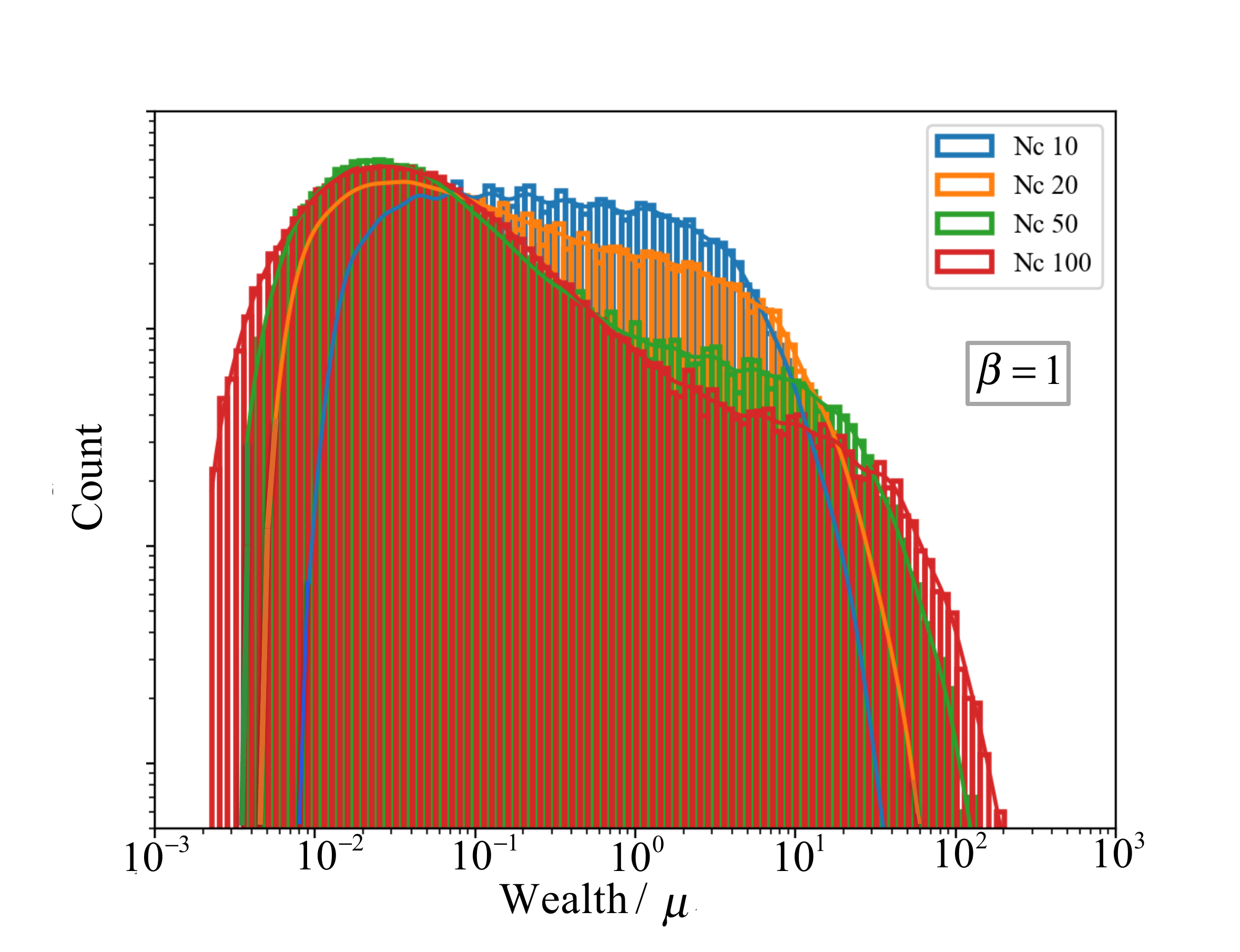}
\includegraphics[width=\columnwidth]{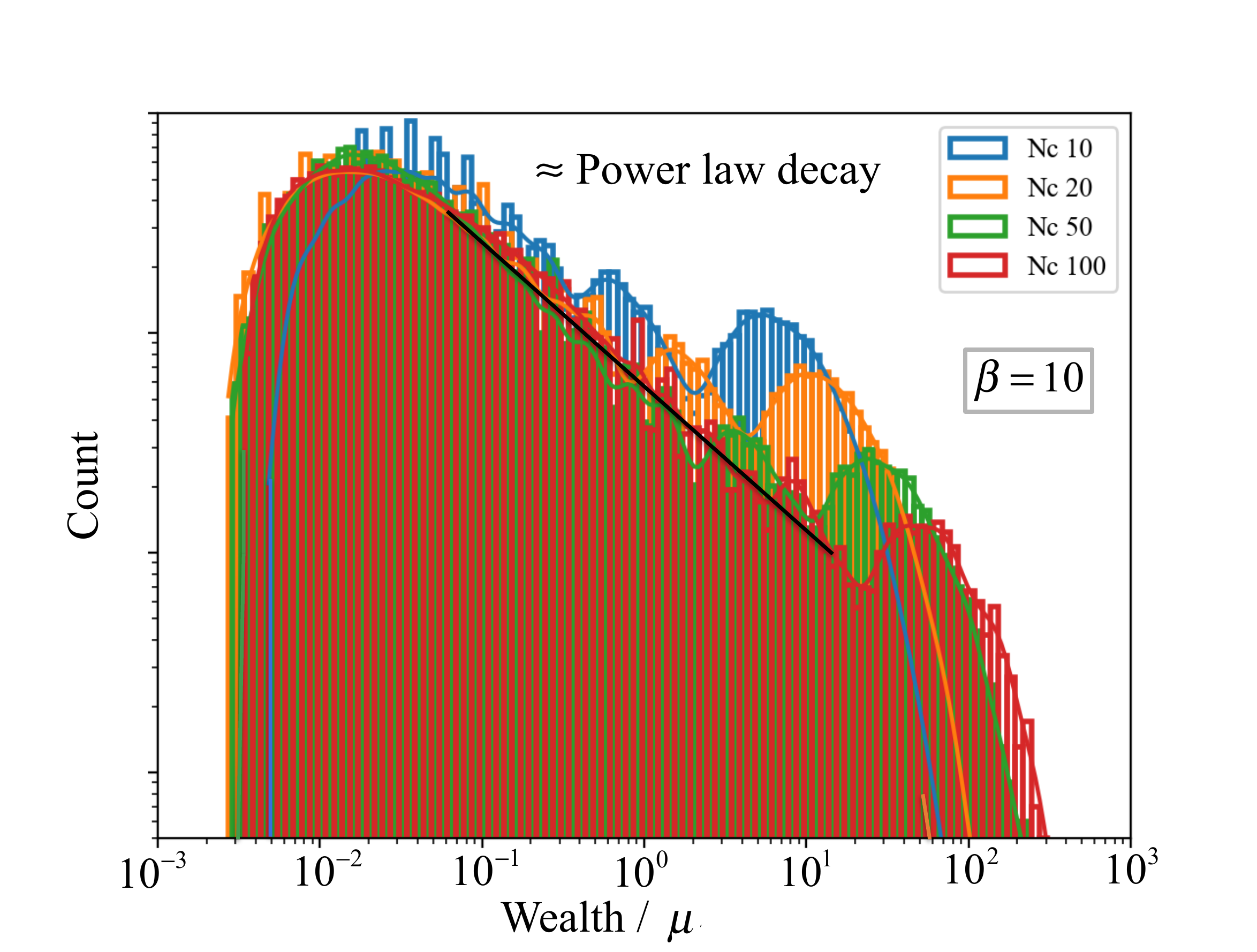}
\caption{Stationary wealth distributions for a society with $N_0=2\times10^5$,
$N_c=\{10,20,50,100\}$, and $\beta=1$ (top) or $\beta=10$
(bottom).~\label{Fig7}}
\end{center}
\end{figure}
Let us now study the dependence on the number of classes $N_c$.
Figure~\ref{Fig7} shows the stationary wealth distribution
for different values of $N_c$ and $\beta=1$ or $\beta=10$. We notice from the
simulations that, the larger the $N_c$ or $\beta$ is, the more generations it
takes for the wealth distribution to converge. While in the previous cases,
ten generations were enough to go from a Gaussian of average $\mu$ and
$\sigma=\mu/10$ to an approximate stationary distribution, for $N_c=100$ and
$\beta=1$ about $40$ generations, and for $N_c=100$ and $\beta=10$ about $60$
generations. 

For $\beta=1$, as $N_c$ increases, the shape of the distribution changes from
an almost flat distribution (with 10 small undulations) for $N_c=10$, to one
with a peak for values below the average wealth $\mu$ and a power-law regime
for values of the wealth around $\mu$. The range of the power-law regime
widens with $N_c$ and the exponent $\approx -2/3$ (slope in the log-log plot)
seems to be independent of $N_c$. For larger wealth, one observes a second
bump that seems to decrease with $N_c$. The Gini coefficient ranges from
$G\approx0.72$ for $N_c=10$ to $G\approx0.93$ for $N_c=100$. The fraction of
the wealth in the top $10\%$ individuals is $56\%$ and $93\%$, respectively. 

For $\beta=10$, the effect of $N_c$ is more pronounced. The Gini coefficient
changes from $G\approx0.87$ for $N_c=10$ to $G\approx0.95$ for $N_c=100$, and
the fraction of the wealth in the top $10\%$ individuals is $84\%$ and $95\%$
respectively. As discussed before for $N_c=3$, the distribution consists of a
sequence of $N_c$ peaks. However, the relative height
between the peaks is such that, as $N_c$ increases, the overall distribution
is consistent with a peak for a value of the wealth below the average and a
power law of the same exponent $\approx -2/3$ as before, independently of
$N_c$, and a bump for large values of the wealth. This bump corresponds to the
wealthiest class. The height of the bump decreases with $N_c$ and its position
moves towards higher values. For $N_c=100$, the power-law regime extends over
three orders of magnitude.

\section{Conclusion}

The societies we studied here were mainly characterized by a certain offspring
probability distribution (Eq. \ref{eq_opd}), accounting for the variable
number of children of each family, and a marriage probability distribution
(Eq. \ref{eq_mpd}) that depends on the society number of different wealth
classes $N_c$ and their level of distinction $\beta$.

In a society without classes, where marriages are random, we found (as in
\cite{blinder1973model}) an egalitarian wealth distribution if all families
have exactly the same number of children. However, wealth inequality emerges
from the moment there is a different offspring distribution, in which families
may have different number of children. Both for the more simple example
$f^o=\{0,1/3,1/3,1/3\}$ and the more realistic choice
$f^o=\{0.1,0.2,0.4,0.2,0.1\}$, we observed in a few ($\approx 4-5$)
generations a rapid evolution to a stationary wealth distribution with
$G\approx 0.3$. The stationary wealth distribution has, for $w>\mu$, an
exponentially decaying (or Boltzmann) law, $f(w)\propto e^{-w/w_c}$, where
$w_c$ is a characteristic wealth.

In societies with classes, stationary wealth distributions were also found for
the representative offspring distribution $f^o=\{0,1/3,1/3,1/3\}$.  The larger
the number ($N_c$) and the distinction between classes ($\beta$), the longer
it takes to attain the stationary distribution, and the larger is its Gini
coefficient. The values we obtained for our model societies reflect the
economical empirical known estimates. While Gini coefficients in developed
countries typically range between about 0.3 and 0.4 for income, they vary from
about 0.5 to 0.9 for wealth \cite{davies2000distribution}.

The stationary distributions may, in certain cases, acquire complex and
undulated shapes. As the distinction between classes increases, we may
observe, in the log-log representation of Fig.~\ref{Fig5}, for a society with
only $N_c=3$ classes, the continuous evolution from a distribution with only
one single peak (for $\beta=0$), to a distribution with $N_c=3$ overlapping peaks
($\beta=10$).  We also observed in the stationary wealth distributions for
other values of $N_c$ the appearance of a number of peaks equal to the society
number of classes.

As the number of classes $N_c$ increases, we observe a power-law regime in the
stationary wealth distribution, for intermediate values of wealth
($10^{-1}\mu<w<10^1\mu$ for $N_c=100$). This power law exists already for a
miscible society (with $\beta=1$) but it extends over three orders of
magnitude for a stricter one ($\beta=10$). The exponent of the power law
remains approximately the same and equal to $\alpha=2/3$, independent of $N_c$
and $\beta$, as soon as these values are large enough. The values of the
exponents observed empirically range in between 1.5 and
3~\cite{atkinson2011top}. As it is well established, the value of power-law
exponents emerging from a non-linear dynamics strongly depend on the spatial
dimension and correlations of the underlying topology~\cite{Dorogovtsev08}.
Here, we have assumed that every individual can, in principle, marry any other
of a different gender (mean field). However, in reality, individuals live in a
time-dependent social network and the likelihood of getting marry also depends
on the effective distance between individuals in such network. How
$\alpha$ depends on the underlying topology is a topic of future work.

\section*{Acknowledgments}
We acknowledge financial support from the Portuguese Foundation for Science
and Technology (FCT) under Contracts no. UIDB/00618/2020 and UIDP/00618/2020. 

\bibliography{wealth}

\end{document}